\documentclass[10pt,letterpaper,twocolumn,prl,showpacs]{revtex4-1}

\usepackage[english]{babel}
\usepackage{amsmath,amssymb}
\usepackage{graphicx}
\usepackage{epstopdf}
\usepackage{color}
\usepackage[usenames,dvipsnames]{xcolor}
\usepackage{hyperref}
\DeclareGraphicsExtensions{.eps,.pdf}

\begin{document}

\title{Large net-normal dispersion Er-doped fibre laser mode-locked \\ with a nonlinear amplifying loop mirror}

\author{Patrick Bowen$^{*}$, Miro Erkintalo, and Neil G. R. Broderick}

\affiliation{[1] The Dodd Walls Centre for Photonic and Quantum Technologies, Department of Physics, University of Auckland, Auckland 1142, New Zealand
\\$^*$Corresponding author: pbow027@aucklanduni.ac.nz}

\begin{abstract}
We report on an environmentally stable, all-PM-fibre, Er-doped, mode-locked laser with a central wavelength of 1550~nm. Significantly, the laser possesses large net-normal dispersion such that its dynamics are comparable to that of an all-normal dispersion fibre laser at 1~$\mu$m with an analogous architecture. The laser is mode-locked with a nonlinear amplifying loop mirror to produce pulses that are externally compressible to 500~fs. Experimental results are in good agreement with numerical simulations. 
\end{abstract}

\maketitle

\noindent  Mode-locked fibre laser systems producing ultrashort ($<$~500~fs) pulses have numerous applications in fields such as  micro-machining, spectroscopy, and nonlinear imaging~\cite{NatureMicroscopy,NatureUltrafastFibreLaser,ThuliumCuttingPolymer,MicroMachiningHeat,OCT,OCT2,CARS,Imaging2,Imaging1}. Because the performance of the end-application is often correlated with the characteristics of the mode-locked pulses, significant research efforts have been invested to devise new and improved laser configurations. To a large extent, the overarching performance and dynamical behaviour of a mode-locked fibre laser is dictated by the dispersion landscape of the cavity. In particular, depending on the cavity dispersion profile, the laser can support one of several distinct pulse regimes, which include e.g. conventional solitons~\cite{solitonAndDisSolitonLaser}, dispersion-managed solitons~\cite{2umLowRepNetZero}, and similaritons~\cite{SolitonSimilariton}.\looseness=-1

Over the last decade, lasers constructed out of all-normal dispersion (ANDi) fibres and components have attracted particular attention~\cite{ChongANDifirst,ChongANDi20nJ,55fsANDiLaserPulses}. This is because such ANDi lasers have been shown to outperform alternative cavity designs in terms of pulse characteristics (e.g. energy, bandwidth), whilst simultaneously simplifying the overall cavity design. These lasers have been dominantly investigated in the 1~$\mu$m emission band of Ytterbium, partly because of the applications in that wavelength region, but also because all standard fibres exhibit normal dispersion at 1~$\mu$m. In many circumstances, however, the applications a laser can service depends critically on the laser’s wavelength of operation~\cite{ThuliumCuttingPolymer,ErbiumGasSensing,ErbiumMachining}. In particular, for certain applications, operation at 1~$\mu$m may not be feasible or optimal, and several studies have accordingly endeavoured to translate ANDi technologies developed in the 1~$\mu$m range to other wavelengths, such as 1.55~$\mu$m (Erbium) or 2~$\mu$m (Thulium)~\cite{ANDiErLaser,2umGumenyuk,NetNormEr10nJ,GCOErlaser,NetNormEr12nJ}.\looseness=-1

The realisation of ANDi-like lasers at the 1.55~$\mu$m emission band of Erbium (and the 2~$\mu$m of Thulium) is hindered by the fact that most conventional fibres exhibit normal dispersion only for wavelengths shorter than 1.3~$\mu$m. Nevertheless, there has been several successful demonstrations of designs that leverage dispersion management so as to achieve ANDi-like performance through a sufficiently large net-normal dispersion, both around 1.55~$\mu$m and 2~$\mu$m~\cite{ANDiErLaser,SolitonSimilariton,liu14selfErbium,ANDiErLaser2,cabasse2009high,AtomicGrapheneEr,GrapeheneErNanotubes,TopologicalInsSA}. Whilst these realisations have demonstrated impressive performance, their choice of mode-locking element (nonlinear polarisation rotation or real saturable absorbers) presents potential deficiencies in terms of environmental stability and long-term performance. In particular, nonlinear polarisation evolution prohibits the use of polarisation-maintaining fibres necessary for environmental stability, whereas many real saturable absorbers suffer from material degradation issues. These challenges could be overcome by using a nonlinear optical amplifying loop mirror (NALM), which has been shown to allow for the mode-locking of environmentally stable ANDi devices in the 1~$\mu$m regime~\cite{ClaudeLaser,ClaudeLargeChirpLaser,RungeDFT,Claude120fsLaser}. However, to the best of our knowledge, there has been no demonstrations of large-net normal devices based on NALMs around the 1.55~$\mu$m emission band of Erbium.\looseness=-1

In this Letter, we report on an Erbium-doped, all-PM-fibre laser that exhibits large net-normal dispersion and is mode-locked with a NALM. The laser has a central wavelength of 1.55~$\mu$m, and it emits pulses that can be externally compressed to 500~fs. We also show results from numerical simulations that are in good agreement with our experimental observations.\looseness=-1

\begin{figure}[t]
\centering
\includegraphics[width=0.8\linewidth]{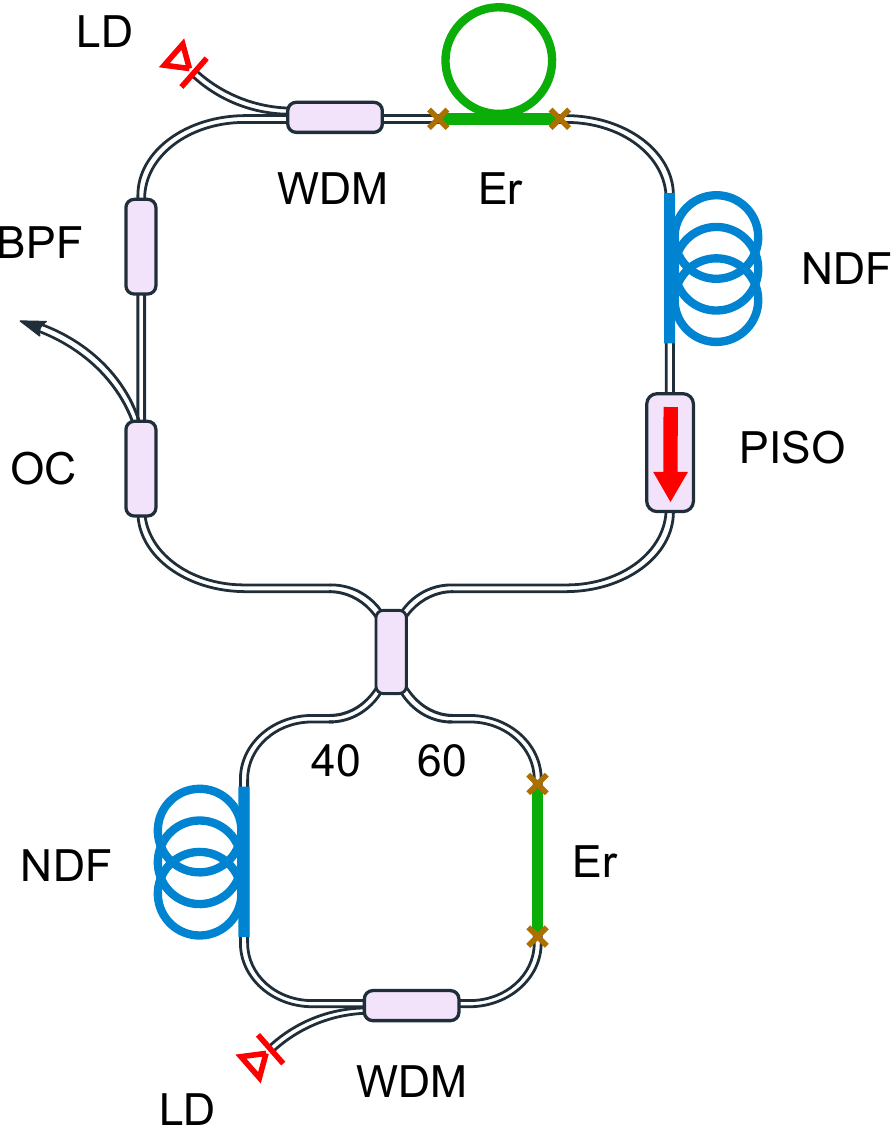}
\caption{Schematic drawing of the studied fibre laser cavity. WDM: wavelength division multiplexer, Er: Erbium-doped active fibre, LD: 980~nm pump laser diode, BPF: 2~nm band-pass filter, PISO: polarising isolator, OC: 70:30 output coupler (30\% out), NDF: passive fibre used in the NALM having normal dispersion. Brown crosses show points of high splice loss as described in the text.}
\label{SchematicGCO}
\end{figure}

The laser cavity under study, shown in Fig.~\ref{SchematicGCO}, is conceptually based on a 1~$\mu$m device reported in~\cite{ClaudeLaser}. It consists of two loops connected with a 2x2 central coupler (60:40), where the main, uni-directional loop contains a 70:30 output coupler (30\% out), 2~nm band-pass filter, 980/1550~nm wavelength division multiplexer (WDM), 1.2~m of Erbium-doped fibre, 10~m of normally dispersive fibre, and a polarising isolator. The other bi-directional loop constitutes the NALM, and contains a further 1.5~m of Erbium-doped fibre, another WDM, and 10~m of normally dispersive fibre. Unlike traditional figure-eight lasers, the design in Fig.~\ref{SchematicGCO} has two separately pumped gain sections, giving greater tunability to the laser.\looseness=-1

Our laser cavity is constructed from off-the-shelf fibre-optic components that use standard single-mode fibre (SMF) with 10.6~$\mu$m mode-field diameter (MFD). Because such SMF possesses large anomalous dispersion at 1.55~$\mu$m, dispersion management is needed to attain operation similar to ANDi lasers. We achieve a large net-normal dispersion cavity by minimising the amount of standard SMF in the cavity, and by making use of speciality optical fibre that has a 4~$\mu$m MFD which allows it to exhibit normal dispersion. More specifically, the cavity contains three different polarisation-maintaining fibres that all have different dispersion values. The Erbium-doped fibre which accounts for 2.7~m of the cavity has a dispersion of $\beta_2^\mathrm{G} = 0.03$~$\textrm{ps}^2/\textrm{m}$; the normally dispersive passive fibre (NDF) which accounts for 20~m has a dispersion of $\beta_2^\mathrm{NDF} = 0.04$~$\textrm{ps}^2/\textrm{m}$; and finally the 2.5~m of standard SMF that makes up the pigtails of various components, has the typical dispersion of $\beta_2^\mathrm{SMF} = -0.024$~$\textrm{ps}^2/\textrm{m}$. This gives a large net cavity dispersion of $\beta_2^\mathrm{net} = 0.82$~$\textrm{ps}^2$ and a length of 25.2~m.\looseness=-1

\begin{figure}[t]
\centering
\includegraphics[width=\linewidth]{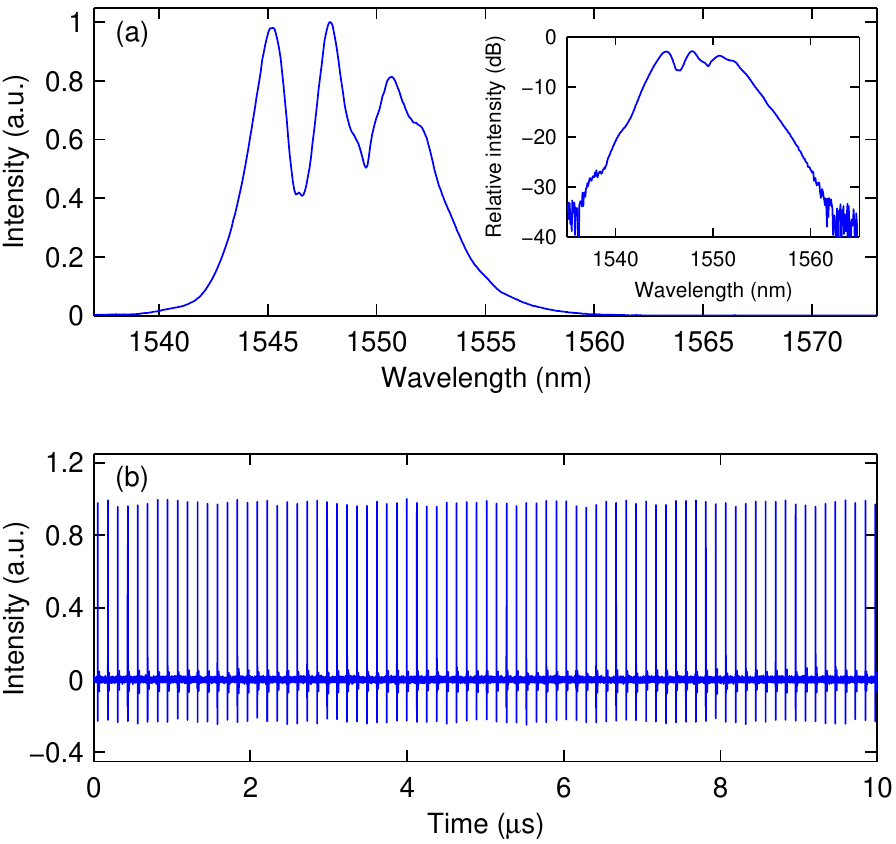}
\caption{Output pulse characteristics of the reported laser cavity. (a) Spectrum in linear scale and inset in log scale, and (b) pulse train recorded on an oscilloscope.}
\label{Figure1}
\end{figure}

The necessity to use fibres with different MFDs presents some challenges for splicing the cavity together. In particular, non-ideal splices between fibres with different MFDs can result in significant losses and parasitic back reflections which prevent the laser from operating. For example, the two different types of passive fibres (CorActive SCF-UN-3/125-25-PM and typical PM1550) have MFDs of 4~$\mu$m and 10.6~$\mu$m, respectively. To achieve a good splice, we use a furnace-based splicing system (Vytran Glass Processing Unit) and a two-step process developed by Wang et al.~\cite{Vytran1}, called ``fire-polishing'', whereby the fibres are weakly connected using a low heat, then continuously heated in a scanning motion across the splice, transforming the MFDs of the fibres to match. After 1--2 minutes of heating, the splice has a measurable transmission of 96\%. Unfortunately, such high efficiencies could only be achieved between the different passive fibres; a reduced transmission (max. 83\%) was achieved when joining the doped fibre (MFD = 4.2~$\mu$m) to the standard SMF (MFD = 10.6~$\mu$m), presumably as diffusion of dopant ions during heating causes the MFD to expand too rapidly~\cite{Vytran1}. In an attempt to increase the light transmitted into the active fibre, an intermediary fibre (SCF-UN-3/125-25-PM) was spliced between the large core PM1550 and the small core active fibre. Since the active fibre shared a similar MFD to the small core passive fibre, a single, short, high power heating process was used to achieve a higher splice transmission (90\%). Unfortunately, when coupled with the second splice transmission (96\%), only a slight increase in total transmission was observed (86\%). As such, this method was not implemented and the brown crosses in Fig.~\ref{SchematicGCO} indicate the points in the cavity where large splice losses occur.\looseness=-1


Despite the comparatively large splice losses, the laser can be mode-locked at a pump power of 60 and 250~mW in the main and the NALM loops, respectively. Similar to the device in \cite{ClaudeLaser}, the laser self-starts with multiple pulses circulating the cavity, but can then be transitioned to the single-pulse regime by reducing the pump powers to 35 mW (main loop) and 55 mW (NALM loop). At these pump powers, the laser produces an average output power of 0.85~mW, corresponding to a pulse energy of 110~pJ at the fundamental repetition rate of 7.9~MHz. The output spectrum of the laser, shown in linear scale in Fig.~\ref{Figure1}(a), was measured with an optical spectrum analyser (Anritsu MS9710) and it can be seen to be centred at 1548.5~nm (corresponding to the centre wavelength of the band-pass filter), with a full width at half maximum (FWHM) of 8.7~nm. To demonstrate that the laser is stably mode-locked, a pulse train is shown in Fig.~\ref{Figure1}(b).\looseness=-1

\begin{figure}[t]
\centering
\includegraphics[width=\linewidth]{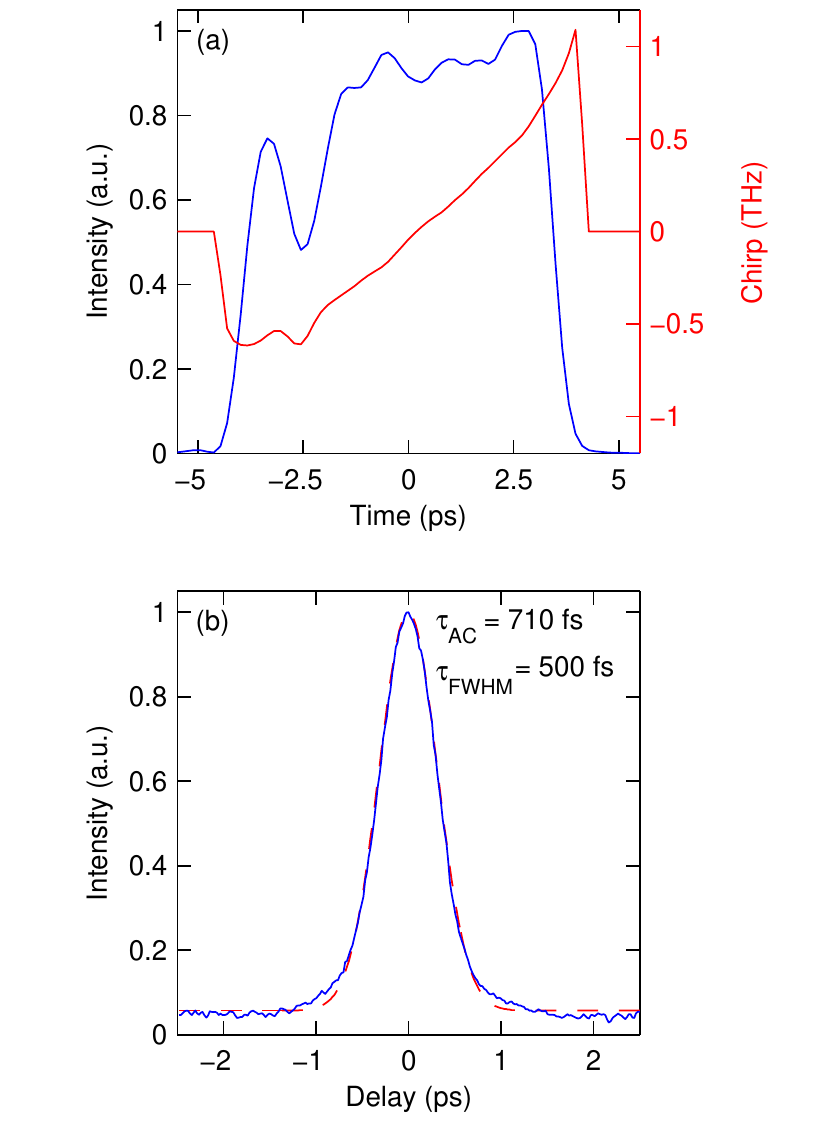}
\caption{Output pulse temporal characteristics for the reported laser cavity. (a) Output envelope and chirp recovered from a measured FROG trace, and (b) autocorrelation trace after external compression (solid blue) and a Gaussian fit (dashed red). $\tau_\mathrm{AC}$ indicates the FWHM of the autocorrelation while $\tau_\mathrm{FWHM}$ is the FWHM of a corresponding Gaussian pulse. }
\label{Figure2}
\end{figure}

To examine the temporal envelope and chirp of the laser output, we used a commercially available frequency resolved-optical gating (FROG) device (Southern Photonics HR150). Figure~\ref{Figure2}(a) shows the temporal envelope and phase as recovered from the measured FROG trace. As can be seen, the pulse has a FWHM duration of 7.3~ps and a nearly-linear chirp. We next attempted to compress the pulse down to ultrashort durations by using an external reflection grating pair with a variable path length. To measure the FWHM of the compressed pulse, we used an autocorrelator (APE pulseCheck) which has a better resolution than our FROG device. Figure~\ref{Figure2}(b) shows that, by adjusting the path length of the compressor, a minimum autocorrelation width of 710~fs can be achieved, which corresponds to a Gaussian pulse shape with a FWHM of 500~fs. Given a spectral width of 8.7~nm, the compressed pulse is then 1.25 times the transform-limit, presumably due to residual nonlinear chirp [see Fig.~\ref{Figure2}(a)] that cannot be compensated for by the grating pair.\looseness=-1

\begin{figure}[t!]
\centering
\includegraphics[width=\linewidth]{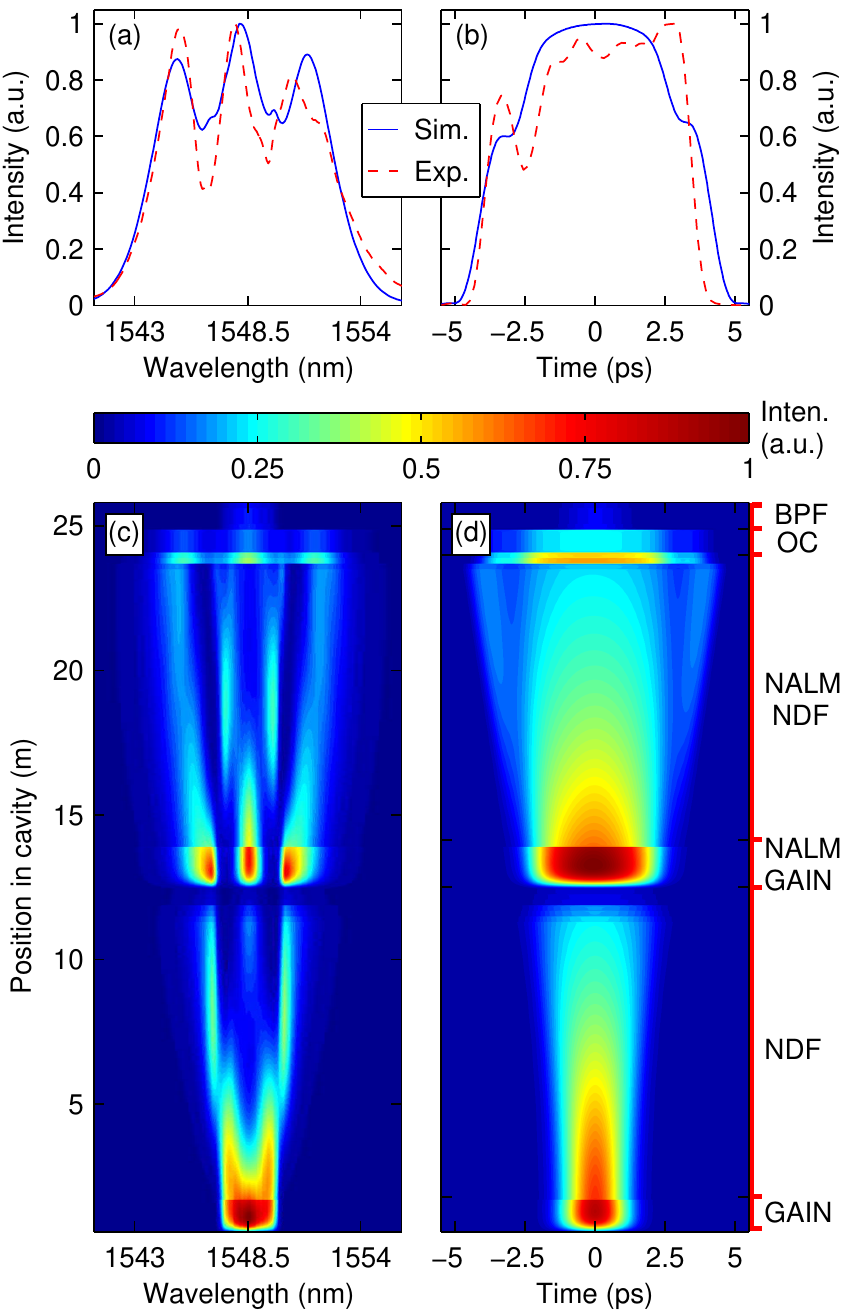}
\caption{Numerical simulations of the laser cavity architecture. (a, b) Blue solid lines show (a) spectral and (b) temporal profiles at the simulated laser output. Dashed red lines indicate experimental results for comparison. Corresponding (c) spectral and (d) temporal evolution of the pulse over one round-trip with relevant sections annotated with labels (e.g. NALM GAIN: active fibre in NALM; NDF: normally dispersive fibre). Note: Only the clockwise pulse propagation in the NALM is shown and the OC section is artificially stretched for clarity.}
\label{FigureSim}
\end{figure}

Comparing the measured laser characteristics with those reported for the 1~$\mu$m device in \cite{ClaudeLaser}, we can see considerable similarity in both the measured spectrum and the recovered time trace. Indeed, both devices are associated with a structured spectrum and a temporal profile with steep edges, as is characteristic for many ANDi lasers. Furthermore, both devices emit pulses with comparatively low sub-nanojoule energies. We note that further developments of the 1~$\mu$m design reported in \cite{ClaudeLaser} has allowed for much larger pulse energies in excess of 10~nJ~\cite{Claude120fsLaser,ClaudeLargeChirpLaser}; however, translating those developments to the current 1.55~$\mu$m device has so far been unsuccessful. For example, when increasing the output coupling from 30\% to 80\%, as in \cite{ClaudeLargeChirpLaser}, we were unable to reach single-pulsing operation. This could be due to the short sections of anomalous dispersion fibre in the cavity, which can lead to detrimental soliton effects, or the increase in nonlinear phase accumulation due to the use of fibres with smaller core diameters. To achieve higher energy pulses with net-normal dispersion, it may be preferable to substitute some amount of normally dispersive fibre for a positively chirped fibre Bragg grating, as in~\cite{2umGumenyuk}. Another effect of using Erbium-doped fibre is the decrease in gain compared to Ytterbium-doped fibre. Specifically, in Erbium-doped fibres, lower doping concentrations must be used due to a tendency towards ion clustering that results in upconversion and an overall decrease in gain efficiency~\cite{ErbiumClustering,ErbiumConcentration}.\looseness=-1


To corroborate our experimental findings, we have also numerically  simulated the operation of our laser. Our simulations use the lumped cavity model that was introduced in~\cite{ClaudeLaser,ClaudeLargeChirpLaser} and described in more details in~\cite{NALMreview}. The model takes into account all the components and different fibre segments in our cavity, and models gain dynamics by using amplifier rate equations. Figures~\ref{FigureSim}(a) and (b) show the spectral and temporal characteristics of the simulated output pulse (solid blue curves), respectively, compared with corresponding experimental results (dashed red lines). As can be seen, the simulations are in very good agreement with experimental findings. The spectral and temporal pulse propagation dynamics are also shown for one full round-trip in Figs.~\ref{FigureSim}(c) and (d), respectively, starting and finishing on either side of the band-pass filter. (For simplicity, only the clockwise propagation of the NALM is shown and the output coupler section is artificially stretched.) An interesting section of the pulse evolution occurs between the end of the normally dispersive fibre in the main loop and the start of the active fibre in the NALM. Between these points, the pulse amplitude is diminished significantly  as it passes several splice transitions and components resulting in an overall loss of 2.4~dB, before being re-amplified to slightly greater than its initial value in the NALM gain, which is very different to the propagation dynamics in~\cite{ClaudeLaser}. Figure~\ref{FigureSim}(c) is also valuable in explaining the structure of the experimentally observed pulse. Specifically, from the regular and continuous spectral broadening, shown in Fig.~\ref{FigureSim}(c), it is clear that the three distinct peaks in the laser output spectrum are simply caused by self-phase modulation. Another point of interest is at the output of the NALM. At this point there is an abrupt change in the pulse shape both spectrally and temporally as the pulse propagating clockwise around the NALM (shown) interferes with the counter-clockwise one (not shown). On the temporal trace in Fig.~\ref{FigureSim}(d), this change is visible as pulse narrowing and an increase in peak power, demonstrating the effectiveness of the NALM as a mode-locking mechanism.\looseness=-1


To summarise, we have presented an all-PM-fibre laser with large-net normal dispersion and a central wavelength of 1550~nm. The laser produces 110~pJ pulses that are externally compressible to 500~fs. This was achieved through dispersion management to produce a linearly chirped externally compressible pulses, and speciality splicing methods to reduce intra-cavity loss. Furthermore, we have shown that the characteristics of the generated pulses agree well with predictions from numerical simulations. For future work,  we seek to leverage our numerical model to explore design modifications, such as the use of fibre Bragg gratings, that could permit increased pulse energies.\looseness=-1

\vspace{12pt}
\noindent \textbf{Funding.} Rutherford Discovery Fellowships of the Royal Society of New Zealand. Marsden Fund of the Royal Society of New Zealand. Ministry of Business, Innovation and Employment, New Zealand.

\newcommand{\enquote}[1]{``#1''}


\begin{thebibliography}{99}

\bibitem{NatureMicroscopy}
C.~Xu and F.~W.~Wise, \enquote{Recent advances in fibre lasers for nonlinear microscopy,} Nature Photonics \textbf{7}, 875--882 (2013).

\bibitem{NatureUltrafastFibreLaser}
M.~E.~Fermann and I.~Hartl, \enquote{Ultrafast fibre lasers,} Nature Photonics \textbf{7}, 868--874 (2013).

\bibitem{MicroMachiningHeat}
S.~Eaton, H.~Zhang, P.~Herman, F.~Yoshino, L.~Shah, J.~Bovatsek, and A.~Arai, \enquote{Heat accumulation effects in femtosecond laser-written waveguides with variable repetition rate,} Optics Express \textbf{13}, 4708--4716 (2005).

\bibitem{ThuliumCuttingPolymer}
G.~L.~Roth, S.~Rung, and R.~Hellmann, \enquote{Welding of transparent polymers using femtosecond laser,} Applied Physics A \textbf{122}, 1--4 (2016).

\bibitem{OCT}
A.~G.~Podoleanu, \enquote{Optical coherence tomography,} The British Journal of Radiology (2014).

\bibitem{OCT2}
M.~T.~Tsai and M.~C.~Chan, \enquote{Simultaneous 0.8, 1.0, and 1.3~$\mu$m multispectral and common-path broadband source for optical coherence tomography,} Optics Letters \textbf{39}, 865--868 (2014).

\bibitem{CARS}
H.~G.~Breunig, M.~Weinigel, R.~B{\"u}ckle, M.~Kellner-H{\"o}fer, J.~Lademann, M.~E.~Darvin, W.~Sterry, and K.~K{\"o}nig, \enquote{Clinical coherent anti-Stokes Raman scattering and multiphoton tomography of human skin with a femtosecond laser and photonic crystal fiber,} Laser Physics Letters \textbf{10}, 025604 (2013).

\bibitem{Imaging2}
C.~L.~Evans and X.~S.~Xie, \enquote{Coherent anti-Stokes Raman scattering microscopy: chemical imaging for biology and medicine,} Annu. Rev. Anal. Chem. \textbf{1}, 883--909 (2008).

\bibitem{Imaging1}
C.~J.~Engelbrecht, R.~S.~Johnston, E.~J.~Seibel, and F.~Helmchen, \enquote{Ultra-compact fiber-optic two-photon microscope for functional fluorescence imaging in vivo,} Optics Express \textbf{16}, 5556--5564 (2008).

\bibitem{solitonAndDisSolitonLaser}
S.~Y.~Choi, H.~Jeong, B.~H.~Hong, F.~Rotermund, and D.~I.~Yeom, \enquote{All-fiber dissipative soliton laser with 10.2 nJ pulse energy using an evanescent field interaction with graphene saturable absorber,} Laser Physics Letters \textbf{11}, 015101 (2013).

\bibitem{2umLowRepNetZero}
L.-M.~Yang, P.~Wan, V.~Protopopov, and J.~Liu, \enquote{2~$\mu$m femtosecond fiber laser at low repetition rate and high pulse energy,} Optics Express \textbf{20}, 5683--5688 (2012).

\bibitem{SolitonSimilariton}
B.~Oktem, C.~{\"U}lg{\"u}d{\"u}r, and F.~O.~Ilday, \enquote{Soliton–similariton fibre laser,} Nature Photonics \textbf{4}, 307--311 (2010).

\bibitem{ChongANDifirst}
A.~Chong, J.~Buckley, W.~Renninger, and F.~Wise, \enquote{All-normal-dispersion femtosecond fiber laser,} Optics Express \textbf{14}, 10095--10100 (2006).

\bibitem{ChongANDi20nJ}
A.~Chong, W.~H.~Renninger, and F.~W.~Wise, \enquote{All-normal-dispersion femtosecond fiber laser with pulse energy above 20nJ,} Optics Letters \textbf{32}, 2408--2410 (2007).

\bibitem{55fsANDiLaserPulses}
W.~H.~Renninger, A.~Chong, and F.~W.~Wise, \enquote{Self-similar pulse evolution in a all-normal-dispersion laser,} Physical Review A \textbf{82}, 021805 (2010).

\bibitem{ErbiumGasSensing}
G.~Whitenett, G.~Stewart, H.~Yu, and B.~Culshaw, \enquote{Investigation of a tuneable mode-locked fiber laser for application to multipoint gas spectroscopy,} Journal of Lightwave Technology \textbf{22}, 813--819 (2004).

\bibitem{ErbiumMachining}
T.~Mizunami and A.~Ehara, \enquote{Femtosecond-pulsed laser micromachining and optical damage by an erbium-doped fiber-laser system,} Microelectronic Engineering \textbf{88}, 2334--2337 (2011).

\bibitem{NetNormEr10nJ}
A.~Ruehl, V.~Kuhn, D.~Wandt, and D.~Kracht, \enquote{Normal dispersion erbium-doped fiber laser with pulse energies above 10 nJ,} Optics Express \textbf{16}, 3130--3135 (2008).

\bibitem{GCOErlaser}
L.~R.~Wang, X.~M.~Liu, and Y.~K.~Gong, \enquote{Giant-chirp oscillator for ultra-large net-normal-dispersion fiber lasers,} Laser Physics Letters \textbf{7}, 63 (2009).

\bibitem{NetNormEr12nJ}
C.~Ouyang, P.~P.~Shum, K.~Wu, J.~H.~Wong, X.~Wu, H.~Q.~Lam, and S.~Aditya, \enquote{Dissipative soliton (12 nJ) from an all-fiber passively mode-locked laser with large normal dispersion,} IEEE Photonics Journal \textbf{3}, 881--887 (2011).

\bibitem{2umGumenyuk}
R.~Gumenyuk, I.~Vartiainen, H.~Tuovinen, and  O.~G.~Okhotnikov, \enquote{Dissipative dispersion-managed soliton 2~$\mu$m thulium/holmium fiber laser,} Optics Letters \textbf{36}, 609--611 (2011).

\bibitem{ANDiErLaser}
N.~B.~Chichkov, K.~Hausmann, D.~Wandt, U.~Morgner, J.~Neumann, and D.~Kracht, \enquote{50 fs pulses from an all-normal dispersion erbium fiber oscillator,} Optics Letters \textbf{35}, 3081--3083 (2010).

\bibitem{liu14selfErbium} 
H.~Liu, Z.~Liu, E.~Lamb, and F.~Wise, \enquote{Self-similar erbium-doped fiber laser with large normal dispersion,} Optics Letters \textbf{39}, 1019--1021 (2014).

\bibitem{ANDiErLaser2}
N.~Chichkov, K.~Hausmann, D.~Wandt, U.~Morgner, J.~Neumann, and D.~Kracht, \enquote{High-power dissipative solitons from an all-normal dispersion erbium fiber oscillator,} Optics Letters \textbf{35}, 2807--2809 (2010).

\bibitem{cabasse2009high}
A.~Cabasse, G.~Martel, and J.~L.~Oudar, \enquote{High power dissipative soliton in an Erbium-doped fiber laser mode-locked with a high modulation depth saturable absorber mirror,} Optics Express \textbf{17}, 9537--9542 (2009).

\bibitem{AtomicGrapheneEr}
H.~Zhang, D.~Y.~Tang, L.~M.~Zhao, Q.~L.~Bao, and K.~P.~Loh, \enquote{Large energy mode locking of an erbium-doped fiber laser with atomic layer graphene,} Optics Express \textbf{17}, 17630--17635 (2009).

\bibitem{GrapeheneErNanotubes}
M.~A.~Chernysheva, A.~A.~Krylov, A.~A.~Ogleznev, N.~R.~Arutyunyan, A.~S.~Pozharov, E.~D.~Obraztsova, and E.~M.~Dianov, \enquote{Transform-limited pulse generation in normal cavity dispersion erbium doped single-walled carbon nanotubes mode-locked fiber ring laser,} Optics Express \textbf{20}, 23994--24001 (2012).

\bibitem{TopologicalInsSA}
J.~Sotor, G.~Sobon, and K.~M.~Abramski, \enquote{Sub-130 fs mode-locked Er-doped fiber laser based on topological insulator,} Optics Express \textbf{22}, 13244--13249 (2014).

\bibitem{ClaudeLaser}
C.~Aguergaray, N.~G.~R.~Broderick, M.~Erkintalo, J.~S.~Chen, and V.~Kruglov, \enquote{Mode-locked femtosecond all-normal all-PM Yb-doped fiber laser using a nonlinear amplifying loop mirror,} Optics Express \textbf{20}, 10545--10551 (2012).

\bibitem{Claude120fsLaser}
C.~Aguergaray, R.~Hawker, A.~F.~J.~Runge, M.~Erkintalo, and N.~G.~R.~Broderick, \enquote{120 fs, 4.2 n{J} pulses from an all-normal-dispersion, polarization-maintaining, fiber laser,} Applied Physics Letters \textbf{103}, 121111 (2013).

\bibitem{ClaudeLargeChirpLaser} 
M.~Erkintalo, C.~Aguergaray, A.~F.~J.~Runge, and N.~G.~R. Broderick, \enquote{Environmentally stable all-{PM} all-fiber giant chirp oscillator,} Optics Express \textbf{20}, 669--674 (2012).

\bibitem{RungeDFT}
A.~F.~Runge, C.~Aguergaray, N.~G.~Broderick, and M.~Erkintalo, \enquote{Coherence and shot-to-shot spectral fluctuations in noise-like ultrafast fiber lasers,} Optics Letters \textbf{38}, 4327--4330 (2013).
\bibitem{Vytran1}
B.~S.~Wang and E.~W.~Mies, \enquote{Advanced topics on fusion splicing of specialty fibers and devices,} Asia-Pacific Optical Communications, 6781 (2007).

\bibitem{ErbiumClustering}
P.~Myslinski, D.~Nguyen, and J.~Chrostowski, \enquote{Effects of concentration on the performance of erbium-doped fiber amplifiers,} Journal of Lightwave Technology \textbf{15}, 112--120 (1997).

\bibitem{ErbiumConcentration}
M.~Shimizu, M.~Yamada, M.~Horiguch, and E.~Sugita, \enquote{Concentration effect on optical amplification characteristics of Er-doped silica single-mode fibers,} Photonics Technology Letters \textbf{2}, 43--45 (1990).

\bibitem{NALMreview}
A.~F.~J.~Runge, C.~Aguergaray, R.~Provo, M.~Erkintalo, and N.~G.~Broderick, \enquote{All-normal dispersion fiber lasers mode-locked with a nonlinear amplifying loop mirror,} Optical Fiber Technology \textbf{20}, 657--665 (2014).

\end{thebibliography}
\end{document}